# Lateral Drop Rebound on Hydrophobic and Chemically Heterogeneous Surface

Tingting Ji, Yongcai Pan, Yufu Shao, Bing He, Binghai Wen[*]

Guangxi Key Lab of Multi-Source Information Mining & Security, Guangxi Normal University, Guilin 541004, China

**Abstract:** A drop rebounding from a hydrophobic and chemically heterogeneous surface is investigated using the multiphase lattice Boltzmann method. The behaviors of drop rebounding are dependent on the degrees of the hydrophobicity and heterogeneity of the surface. When the surface is homogeneous, the drop rebounds vertically and the height is getting higher and higher with increases of the surface hydrophobicity. When the surface consists of two different hydrophobic surfaces, the drop rebounds laterally towards the low hydrophobic side. The asymmetrical rebounding is because the unbalanced Young's force exerted on the contact line by the high hydrophobic side is greater than that by the low hydrophobic surface. A set of contours of momentum distribution illustrate the dynamic process of drop spreading, shrinking and rebounding. This work promotes the understanding of the rebound mechanism of a drop impacting the surface and also provides a guiding strategy for precisely controlling the lateral behavior of rebounding drops by hydrophobic degrees and heterogeneous surfaces.

[*]Corresponding author. Email: oceanwen@gxnu.edu.cn.

# 1. Introduction

Drop impact on solid surfaces can be found in a variety of natural phenomena and industrial applications. In nature, successive water drops hollow a stone, and splashing drops produce aerosols, cause erosion, make sounds, and bring the smell of the earth during rain.[1, 2] In industry, drop impacting plays an increasing role in surface cooling,[3] spray coating,[4] inkjet printing [5], fuel combustion, and pesticide applications.[6, 7]

Researchers used a variety of simulation methods and experiments to study the behavior of drops impacting on solid surfaces, which involves important practical applications. Base on experimental and theoretical analysis, Bird *et al.*[8] proved that the fluid dynamics of the drop can be changed by designing a hydrophobic surface with a morphology, thereby reducing the time of the drop on the surface. By engineering textures on the surface, Malouin *et al.*[9] showed that drop rebounding on textured surfaces is significantly influenced by the uniformity of the surface roughness. Antonini *et al.*[10] investigated spreading of millimetric water drops impact on hydrophilic to superhydrophobic surfaces, to evaluate the effect of surface wettability, and also found that drop spreading scaling time changes with wettability. Yao *et al.*[11] proved that the higher the hydrophobicity is, the larger the contact angle will be, and the contact angle hysteresis increases with decreasing of the micro-pillar spacing by designing a hybrid surface consisting of an array of hydrophobic and hydrophilic sites. Weisensee *et al.*[12] studied that surface elasticity also affects drop impact, where a drop impacting an elastic superhydrophobic surface can lead to a two-fold reduction in contact time compared to equivalent rigid surfaces. Li *et al.*[13] showed the translational motion of the impacting drop can be converted to gyrate through heterogeneous surface wettability regulation, and it is also analyzed that the gyration behavior is enabled by the synergetic effect of the asymmetric pinning forces originated from surface heterogeneity and the excess surface energy of the spreading drop after impact. Hao *et al.*[14] used high-speed photography to study the effects of oblique drop on the smooth surfaces under different dip angle and pressure, and the results showed that increasing dip angles or decreasing environmental pressure could completely inhibit drop spatter. Han *et al.*[15] proposed that the asymmetric expansion on the surface is driven by either the Coanda effect or inertia depending on the ratio of the drop diameter to the curvature diameter. Lu *et al.*[16] studied the phenomenon of drops hitting the homogeneous structure superhydrophobic surface with the gradient non-wettability, and found that fallen drop mainly deviates toward weaker regions because of water repellency. Through further analysis, they discovered that the asymmetric mechanic behavior is mainly caused by the unbalanced retraction force between both ends of the impact drop.

In recent years, the numerical simulation method based on lattice Boltzmann has attracted people's attention when studying the impact of the drop on the surface of objects. Zhang et al.[17] applied the lattice Boltzmann simulation method to study the effect of the drop on superhydrophobic surfaces with a gradient of wettability. It is proved that the drop rebound toward both directions on textured surfaces with a wettability gradient. Yuan et al.[18] applied the lattice Boltzmann simulation method to simulate the dynamic behavior of the drop impacting on surface with randomly distributed rough micro-structures. The results show that the distribution of surface asperity has a significant effect on the shape of drop spreading. Peng et al.[19] used the multiphase lattice Boltzmann method to simulate the dynamic evolution of drop and the velocity distribution inside the drop during coalescence, and the jumping height caused by drop coalescence on the high hydrophobic surface is predicted based on the energy conservation method. Raman et al.[20] investigated influence of receding contact angle and wettability gradient, and analyzed the coupled dynamics of the impacting drop on the solid surface. It shows that the receding contact angle is the key wetting parameter that determines the final physical outcome of the impact process. Hoseinpour et al.[21] used the lattice Boltzmann method to simulate the collision of a drop with inclined dry walls under gravitational force, indicating that higher density ratio reinforced gravitational force and resulted in more deformation of the drop.

Although many achievements have been made, there is still a lack of detailed analysis on the quantitative process and internal mechanism of the drop rebounding on the chemically heterogeneous surface. In this paper, we used the multiphase lattice Boltzmann method driven by a chemical potential[22-24] to quantitatively analyze the rebounding behavior of the drop. First, introduce the model and methods used in the simulation. On this basis, a series of rebounding behaviors of drops after impacting the chemical surface are studied and analyzed. Subsequently, the dynamic mechanism of the drop from contacting to leaving the hydrophobic and chemically heterogeneous surface is analyzed. Finally, we conclude with a brief summary of this work.

## 2. Multiphase Lattice Boltzmann Method

### 2.1. Multiple-Relaxation-Time Lattice Boltzmann Model

The lattice Boltzmann equation (LBE) is fully discretized in time, space, and velocity. Its multiple-relaxation-time (MRT) version can be concisely expressed as

$$f_i(\mathbf{x}+\mathbf{e}_i,t+1) - f_i(\mathbf{x},t) = -\mathbf{M}^{-1} \cdot \mathbf{S} \cdot [\mathbf{m} - \mathbf{m}^{(eq)}] + F_i \tag{0}$$

Where $f_i(\mathbf{x},t)$ is the particle distribution function at time $t$ and lattice site $\mathbf{x}$, moving along the direction defined by the discrete velocity vector $\mathbf{e}_i$ with $i = 0, ..., N$, $F_i$ is the body force term, $\mathbf{m}$ and $\mathbf{m}^{(eq)}$ represent the velocity moments of the distribution functions and their equilibria, $\mathbf{M}$ is a transformation matrix which linearly transforms the distribution functions to the velocity moments, $\mathbf{m} = \mathbf{M} \cdot \mathbf{f}$, where $\mathbf{f} = (f_0, f_1, ..., f_N)$. $\mathbf{S}$ is a diagonal matrix of relaxation times and is expressed as $\mathrm{diag}(s_\rho, s_e, s_\varepsilon, s_j, s_q, s_j, s_q, s_v, s_v)$ for the two-dimensional nine-velocity (D2Q9) model. In this paper, the relaxation times are given by $s_\rho = s_j = 1$, $s_e = 1.64$, $s_\varepsilon = 1.54$, $s_q = 1.7$, $s_v = 1/\tau$, and $\tau = 0.7$.

The mass density and momentum of fluid at each lattice node are defined by

$$\rho = \sum_{i=0}^{8} f_i, \qquad \rho \mathbf{u} = \sum_{i=0}^{8} \mathbf{e}_i f_i. \tag{0}$$

The external force is incorporated into LBE by the exact difference method,[25] in which the body force term $F_i$ is simply equal to the difference of the equilibrium distribution functions before and after the nonideal force acting on the fluid during a time step.

$$F_i = f_i^{(eq)}(\rho, \mathbf{u} + \delta\mathbf{u}) - f_i^{(eq)}(\rho, \mathbf{u}), \tag{0}$$

where $\delta\mathbf{u} = \delta t \mathbf{F}/\rho$. Correspondingly, the macroscopic fluid velocity is redefined by the average momentum before and after the collision: $\mathbf{v} = \mathbf{u} + \delta t \mathbf{F}/(2\rho)$.

## 2.2. Chemical-potential Multiphase Model

Following the classical capillarity theory of van der Waals, the free energy functional within a gradient-squared approximation is[26-28]

$$\Psi = \int [\psi(\rho) + \frac{\kappa}{2}|\nabla\rho|^2]d\bm{x},  \tag{0}$$

where $\rho$ is the local density, $\kappa$ is the surface tension coefficient, $\psi$ is the bulk free energy density at a given temperature, and the square of the gradient term gives the free energy contribution from density gradients in an inhomogeneous system. The chemical potential can be derived from the free energy density functional: [26, 29, 30]

$$\mu = \psi'(\rho) - \kappa\nabla^2\rho. \tag{0}$$

The nonlocal pressure is related to free energy by

$$p = p_0 - \kappa\rho\nabla^2\rho - \frac{\kappa}{2}|\nabla\rho|^2, \tag{0}$$

with the general EOS defined by the free energy density

$$p_0 = \rho\psi'(\rho) - \psi(\rho). \tag{0}$$

The thermodynamic pressure tensor of non-uniform fluids contains the non-diagonal terms

$$P_{\alpha\beta} = [p_0 - \kappa\rho\nabla^2\rho - \frac{\kappa}{2}(\nabla\rho)^2]\delta_{\alpha\beta} + \kappa\frac{\partial\rho}{\partial x_\alpha}\frac{\partial\rho}{\partial x_\beta} \tag{0}$$

where $\delta_{\alpha\beta}$ is the Kronecker delta function. With respect to the ideal gas, the nonideal force can be evaluated by the chemical potential [24]:

$$\bm{F} = -\rho\nabla\mu + c_s^2\nabla\rho, \tag{0}$$

where $c_s$ is the sound speed.

The above nonideal force was considered to be evaluated in the momentum space. To improve the stability of the multiphase at the large density ratio, a proportional coefficient $k$ was introduced to decouple the dimension unit of the length between the momentum space and the mesh space,[31]

$$\hat{\delta x} = k \delta x. \tag{0}$$

Here, the symbols of the quantities with a length dimension are marked with a superscript, such as for the lattice length, velocity, and nonideal force. The time, density, and temperature are considered to be independent of the length, so they keep the same symbols and values in the two spaces. With the dimensional analyses, this proportional relationship creates the following transformations,

$$\hat{\mu} = k^2 \mu, \tag{0}$$

$$\hat{\boldsymbol{F}} = k \boldsymbol{F}, \tag{0}$$

$$\hat{\psi}(\rho) = k^2 \psi(\rho), \quad \hat{\psi}'(\rho) = k^2 \psi'(\rho), \tag{0}$$

$$\hat{\nabla} = k^{-1} \nabla, \quad \hat{\nabla}^2 = k^{-2} \nabla^2, \tag{0}$$

the LBE of the present model is evolving in the mesh space (the computational mesh). Because the universal gas constant, attraction parameter, and volume correction have complex dimensions, we limit the equations involving these parameters, such as EOS and free energy density, are calculated in the momentum space, and then they are transformed into the mesh space by the proportional coefficient. Thus, the chemical potential and the nonideal force in the mesh space are evaluated by

$$\hat{\mu} = k^2 \psi'(\rho) - \hat{\kappa} \hat{\nabla}^2 \rho, \tag{0}$$

$$\hat{\boldsymbol{F}} = -\rho \hat{\nabla} \mu + \hat{c}_s^2 \hat{\nabla} \rho. \tag{0}$$

When the coefficient is less than 1, the steep transition region in the momentum space is stretched into a gentler curve in the mesh space. In terms of the result, this is similar to the mesh refinement; however, the present model creates the dimensional transformation between the momentum and mesh spaces with no loss of accuracy. To compute the gradients with a higher accuracy, we further apply the fourth-order accurate central difference method

$$u_i' = \frac{2}{3}(u_{i+1} - u_{i-1}) - \frac{1}{12}(u_{i+2} - u_{i-2}), \tag{0}$$

where $u$ represents the density or chemical potential at each lattice node.

The Peng–Robinson (PR) EOS is used to model the present water-vapor system. Its equation and the relevant chemical potential are written as[24, 31]

$$p_0 = \frac{\rho RT}{1-b\rho} - \frac{a\alpha(T)\rho^2}{1+2b\rho-b^2\rho^2},\tag{0}$$

$$\mu = RT\ln\frac{\rho}{1-b\rho} - \frac{a\alpha(T)}{2\sqrt{2}b}\ln\frac{\sqrt{2}-1+b\rho}{\sqrt{2}+1-b\rho} + \frac{RT}{1-b\rho} - \frac{a\alpha(T)\rho}{1+2b\rho-b^2\rho^2} - \kappa\nabla^2\rho,\tag{0}$$

where the universal gas constant is $R = 1$, the attraction parameter is $a = 2/49$ and volume correction is $b = 2/21$. To relate the numerical results to real physical properties, the reduced variables are defined as $T_r = T/T_c$ and $\rho_r = \rho/\rho_c$, where $T_c$ is the critical temperature and $\rho_c$ is the critical density, and they take $T_c = 0.072919$ and $\rho_c = 2.65730416$ for the PR EOS. The temperature function is $\alpha(T) = [1+(0.37464+1.54226\omega-0.26992\omega^2)\times(1-\sqrt{T/T_c})]^2$ and the acentric factor $\omega$ is 0.344 for water. Obviously, the above computations are all within thermodynamics, the phase transition theoretically satisfies thermodynamics and Galilean invariance, and the satisfactions have been confirmed by static and dynamic numerical simulations and achieve extremely large density ratios.[24, 31]

### 2.3. Chemical-potential Boundary Condition and Contact angle measurement

The chemical potential of a solid surface relates to the surface free energy and can effectively express the surface wettability. Since the present multiphase model is driven by a chemical potential, it is simple and quiet nature to implement the chemical-potential boundary condition[23, 24]. The flat surface of the solid locates on a row of lattice nodes. These nodes are treated as fluid nodes and their distribution functions still collide and stream. The bounce-back boundary condition is applied to mimic those distribution functions from the solid. The solid nodes are assigned a specific chemical potential, which indicates the wetting property of the solid surface. It influences the nonideal force on the interfacial fluid nodes by Eq. (17) and then reflects the interaction between fluid and the solid surface. In order to evaluate the density gradient in Eq. (17) on the interfacial fluid nodes, the densities on the neighbor solid nodes have to be estimated. A simple weighted average scheme based on the neighbor fluid nodes is applied[23]

$$\rho_s(x) = \frac{2}{3}\rho_f(x) + \frac{1}{6}\rho_f(x-1) + \frac{1}{6}\rho_f(x+1),  \tag{0}$$

Where $\rho_s$ represents the solid nodes and $\rho_f$ represents the neighbor fluid node. Using the chemical-potential boundary condition, the contact angle increases almost linearly with the assigned chemical potential of the solid surface.[24] A linear-tunable contact angle is very convenient to adjust the wettability of the solid surface in numerical simulations. During the drop impact on the solid surface, the dynamic contact angles are on-the-spot measured by a geometry-based scheme. Base on the real-time contact angle, the mechanical analyses near the contact line, namely the interactions between the drop and surface, are easily implemented. Figure 1 shows the schematic diagram of contact angle measurement and wetting of chemical potential boundary conditions.

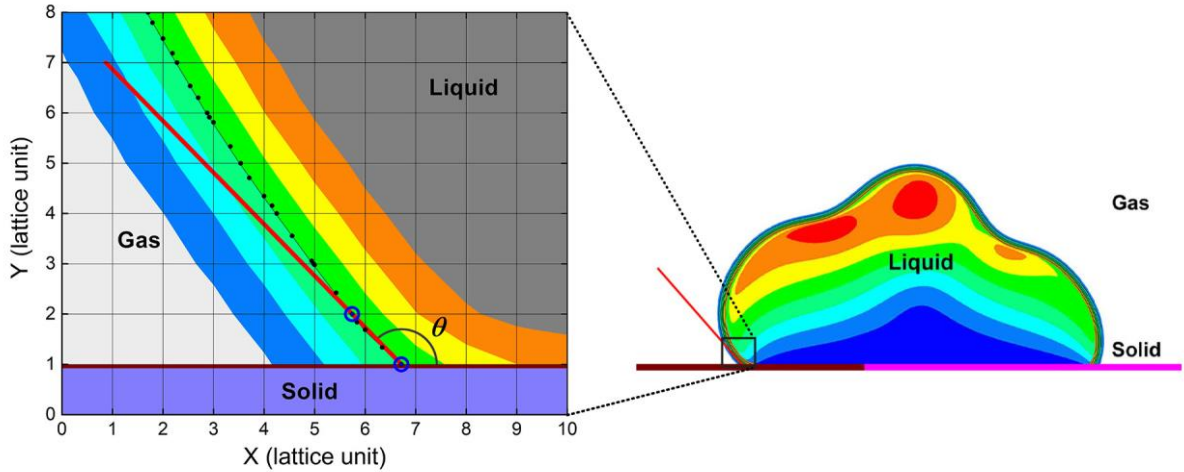

*Figure 1. Schematic diagram of contact angle measurement and wetting of chemical potential boundary conditions.*

### 3. Simulations and Discussion

The chemical-potential multiphase model is used to quantitatively investigate the drop rebound on a chemically heterogeneous solid surface. The computational domain is a rectangle with the width 1200 lattice units and the height 600 lattice units. The circular drop has the radius 75 lattice units and is placed at the lattice node (600, 125), moving down to the solid surface with the initial velocity 0.06. The macroscopic size and density of the drop are

0.25 mm and 1mg/mm$^3$, respectively. The gravity acceleration is 9800 mm/s$^2$. The temperature is set as $T_r = 0.6$, at which the liquid-gas density ratio is near to that of normal temperature.

The solid surfaces used in the present study are always plane and hydrophobic (*i.e.*, $\theta >$ 110°). Four schemes are applied to modify the solid surface with different wettability. The scheme I is a serial of homogeneous surfaces with increasing hydrophobicity. The solid surfaces of the scheme II ~ IV consist of two parts with different contact angles, namely they are chemically heterogeneous surfaces. In the scheme II, the right half is fixed low hydrophobic, while the left half has increasing contact angles. In the scheme III, the left half is fixed high hydrophobic, while the right half has decreasing contact angles. The scheme IV includes two hydrophobic parts with the constant difference of contact angles. Furthermore, we draw the phase diagram of the drop bounce and make mechanical analyses in the drop impact process.

### 3.1. Homogeneous surface (Scheme I)

The solid surfaces in the scheme I have single wettability, namely homogeneous chemical potential. The contact angles grow gradually from 130° to 170° with the interval of 10°. Since the interactions between the drop and surface are symmetrical on the two sides during the drop impacts the surface, the drop can only bounce vertically. Trajectories of the drop centroid movements with the time are drawn in Figure 2(a). The dash lines represent that the drops are touching the surface, while the solid lines indicate that the drops have left the surface. When the contact angle reaches 150°, the phenomenon of twice rebounding occurs. When the contact angle reaches 170°, the drop can rebound three times. In the process of a drop impacting and bouncing, its kinetic energy is always reduced because of the viscous dissipation, and this limits the bouncing time in a small range and the rebounding height is lower and lower. Observing the drop trajectories, the mass centers of the drops leaving the surface are all higher than those of falling back to the surface. Because of the adhesion between the drop and surface, the bouncing drop is elongated, and then the mass center is at a high position. After the drop leaves the surface, the surface tension shrinks it towards a circular shape. Therefore, when the drop falls to touch the surface again, its mass center keeps at a low position.

Figure 2(b) shows the rebounding height and the flying time of the drop impacting on the different homogeneous surfaces. With the enhancement of the surface hydrophobicity, the adhesion that the surface exerts on the drop gets weaker and weaker. Thus, the more kinetic energy is retained to support the drop rebounding, the drop rebounds higher and higher, and the flying time of the first rebounding is longer and longer.

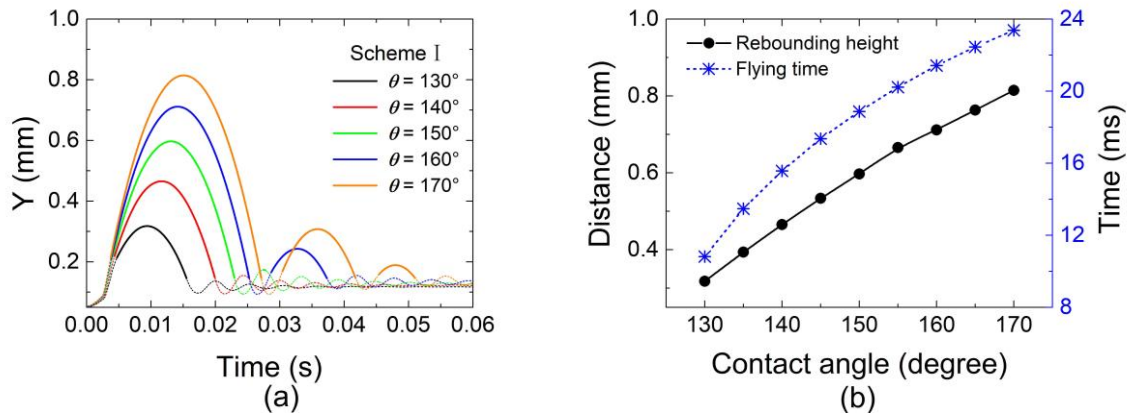

*Figure 2. The drop impacting on homogeneous surfaces. (a) Trajectories of the drop centroid movements after impacting. The solid line represents the drop leaving the surface of the object. (b) The rebounding height and the flying time of the drop on homogeneous surfaces.*

### 3.2. Fixed Low Hydrophobic Side (Scheme II)

The scheme II consists of two parts with different hydrophobicity. The contact angle of the right half is fixed to be low hydrophobic (120°), while those of the left half increase gradually from 130° to 170° with the interval of 10°. Since the surface is asymmetrical, the drop moves towards the low hydrophobic part and the lateral rebounding occurs. The trajectories of the bouncing drops are drawn in Figure 3(a). With the increase of the contact angles of the left part, the lateral distance of the drop rebounding is farther and farther, whereas the rebounding height changes little. The main reason is that as the left-right angle difference increases, the left-right asymmetry force increases. In this scheme, we can't observe the phenomenon of multiple rebounding of the drop. This is mainly due to the fact that on the side of the fixed low hydrophobic surface (120°) on the right, the surface adhesion is too large, and the drop dissipates too much energy on the breakaway surface, and there is no excess energy left to support the drop to rebound repeatedly. The trends are more clearly drawn in Figure 3(b). The flying time of the drop always extends; however, the growth rate before the contact angle 145 °is obviously larger than that of the rest part.

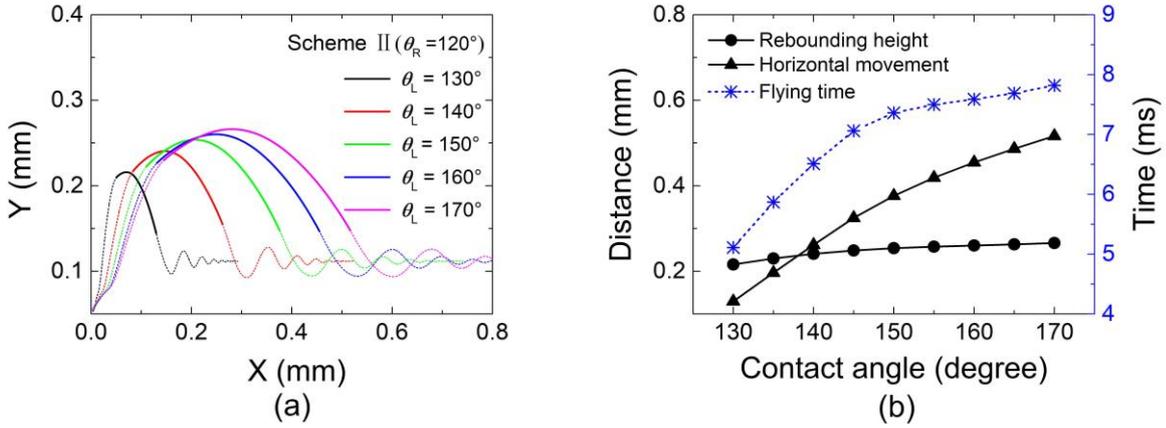

*Figure 3. The drop impacting on heterogeneous surfaces. (a)Trajectories of the drop centroid movements after impacting. The solid line represents the drop leaving the surface of the object. (b) The rebounding height, lateral movement distance and the flying time of the drop on different heterogeneous surfaces.*

### 3.3. Fixed High Hydrophobic Side (Scheme III)

The scheme III sets the left part to be constant and superhydrophobic (CA=170º), while the right part has a serial hydrophobicity from 120 °to 160 °with the interval of 10 °. We can see clearly from Figure 4(a) that with the hydrophobic enhancement of the right part, the rebounding height of the drop gradually increases. However, the lateral movement distance changes neither linearly nor monotonously. Figure 4(b) clearly shows that the lateral movement distance increases at first and then decreases with the increase of the right contact angle. The launching angle is near 45 °when the right part is around 130 °, and the lateral movement distance reaches the maximum. After that, the vertical kinetic energy of the drop is greater than the horizontal kinetic energy. The kinetic energy in the vertical direction of the drop gradually dominates, which promotes the increase of the rebounding height of the drop, and the lateral rebounding distance begins to decrease. However, the flying time always increases monotonously.

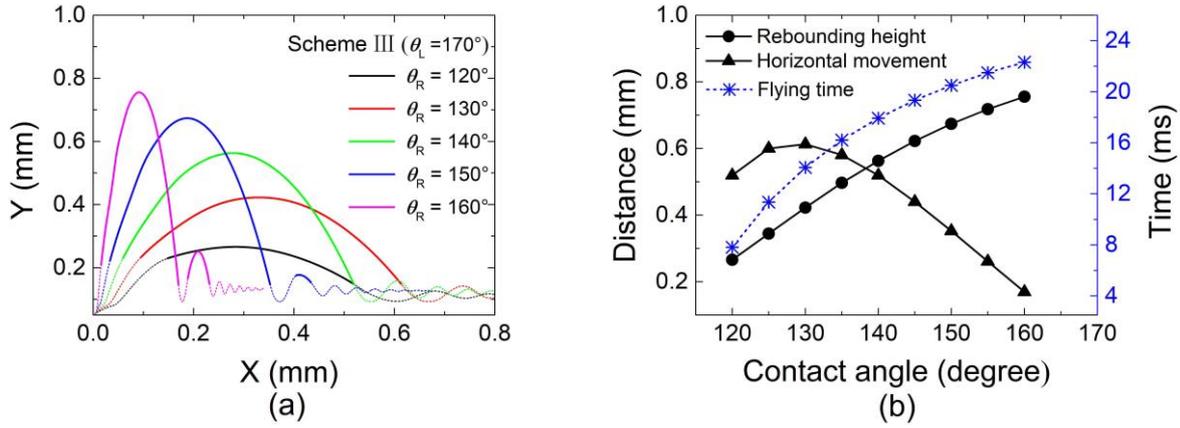

*Figure 4. The drop impacting on heterogeneous surfaces. (a)Trajectories of the drop centroid movements after impacting. The solid line represents the drop leaving the surface of the object. (b) The rebounding height, lateral movement distance and the flying time of the drop on different heterogeneous surfaces.*

### 3.4. Fixed Difference of Contact Angles (Scheme IV)

In scheme IV, the hydrophobicity of the left and right parts is changed simultaneously, but their difference keeps the same. Thus, the configurations of the contact angle are 140º−120º, 150º−130º, 160º−140º, and 170º−150º, and the difference is 20º. As shown in Figure 5(a), we can clearly see that as the left and right angles increase at the same time, the corresponding hydrophobicity continues to increase; the rebounding height gradually increases; however, the lateral distances of the drop rebounding first increases and then decreases. Moreover, the lateral distance of the 160º−140º combined surface reaches its maximum. The flying time of the drop keeps increasing. This trend can be clearly seen from Figure 5(b). This scheme further found that when the left and right angles differences are equal, the rebounding trajectory of the drop is related to the different hydrophobicity of the surface of the object. The simulation analysis of this phenomenon may help to control the placement and trajectory of the drop after impacting.

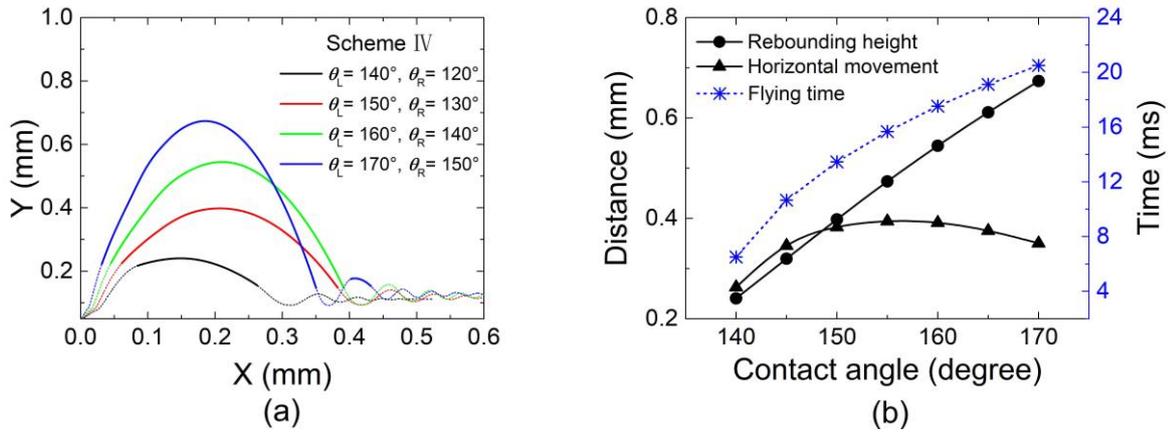

*Figure 5. The drop impacting on heterogeneous surfaces. (a)Trajectories of the drop centroid movements after impacting. The solid line represents the drop leaving the surface of the object. (b) The rebounding height, lateral movement distance and the flying time of the drop on different heterogeneous surfaces.*

### 3.5. Phase diagram of bouncing time

From the above simulation results, we know that when the surface is hydrophobic enough, the drop will rebound twice or more. Further, a series of simulation experiments are implemented to investigate the rebound times and the results are shown in Figure 6. The phase diagram shows that the occurrence of drop lateral rebound requests one of the contact angles at least 115°, meanwhile the other is at least 155°. Except for the homogeneous surface with 120°, all of the homogeneous and heterogeneous surfaces with 120° or larger contact angles can produce the drop rebounding phenomena. The one rebounding is up to 150° contact angle. More than 150°, the drop rebounds many times. In the range of 150° to 165°, the rebounding times are two. When the hydrophobicity has a contact angle of 170°, the rebounding times are three. Especially, the homogenous surface with the contact angle 175° produces four times of drop rebounding. The above four schemes have been labeled in the phase diagram.

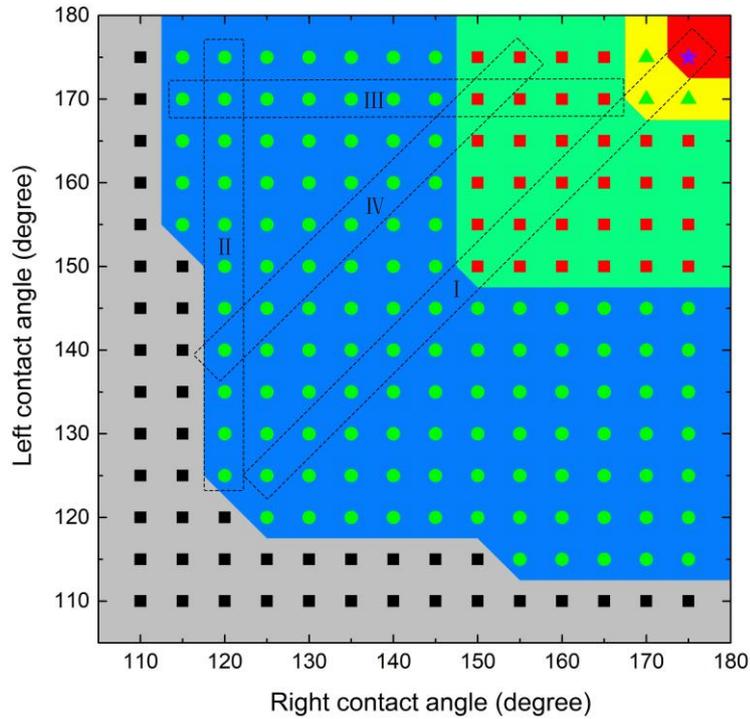

*Figure 6. Phase diagram: Shown on the diagram are rebounding times of the drop on heterogeneous surface. The black squares represent that the drops cannot rebound off, the green circles represent the drops rebound once, the red squares represent the drops rebound 2 times, the green triangles represent the drop rebound 3 times, and the purple five-pointed stars represent the drop rebound 4 times. The ones in the dashed frame are those implemented by the above four schemes.*

### 3.6. Momentum distribution inside the drop

To further understand the inherent mechanisms of the drop, we study the momentum change inside the drop. Figure 7($a_1$−$a_6$) shows the change diagram of the momentum module when the drop falls on the surface with homogeneous and contact angle is 150°, in which ($a_1$−$a_6$) represents the change of the momentum module of the drop at different moments. With the evolution of the simulation time step, when a drop contacts a solid surface, the dynamic change of the drop often produces a series of behaviors. First of all, the drop expands rapidly, using kinetic energy to increase the surface energy of the gas and liquid. Due to the influence of various resistances during the diffusion process of the drop, the diffusion speed of the drop gradually decreases; when the expansion radius of the drop reaches the maximum, it begins to contract in the opposite direction. When the translational energy is completely converted into surface energy, the drop begins to bounce off the surface of the object. It can be seen from Figure 7($a_1$−$a_6$) that the drop is symmetric throughout the spread and retraction process, and its internal momentum module change is also symmetrically distributed. The drop spreads and retracts on the substrate at a constant contact angle, and its center of mass doesn't deviate from the central position of the substrate.

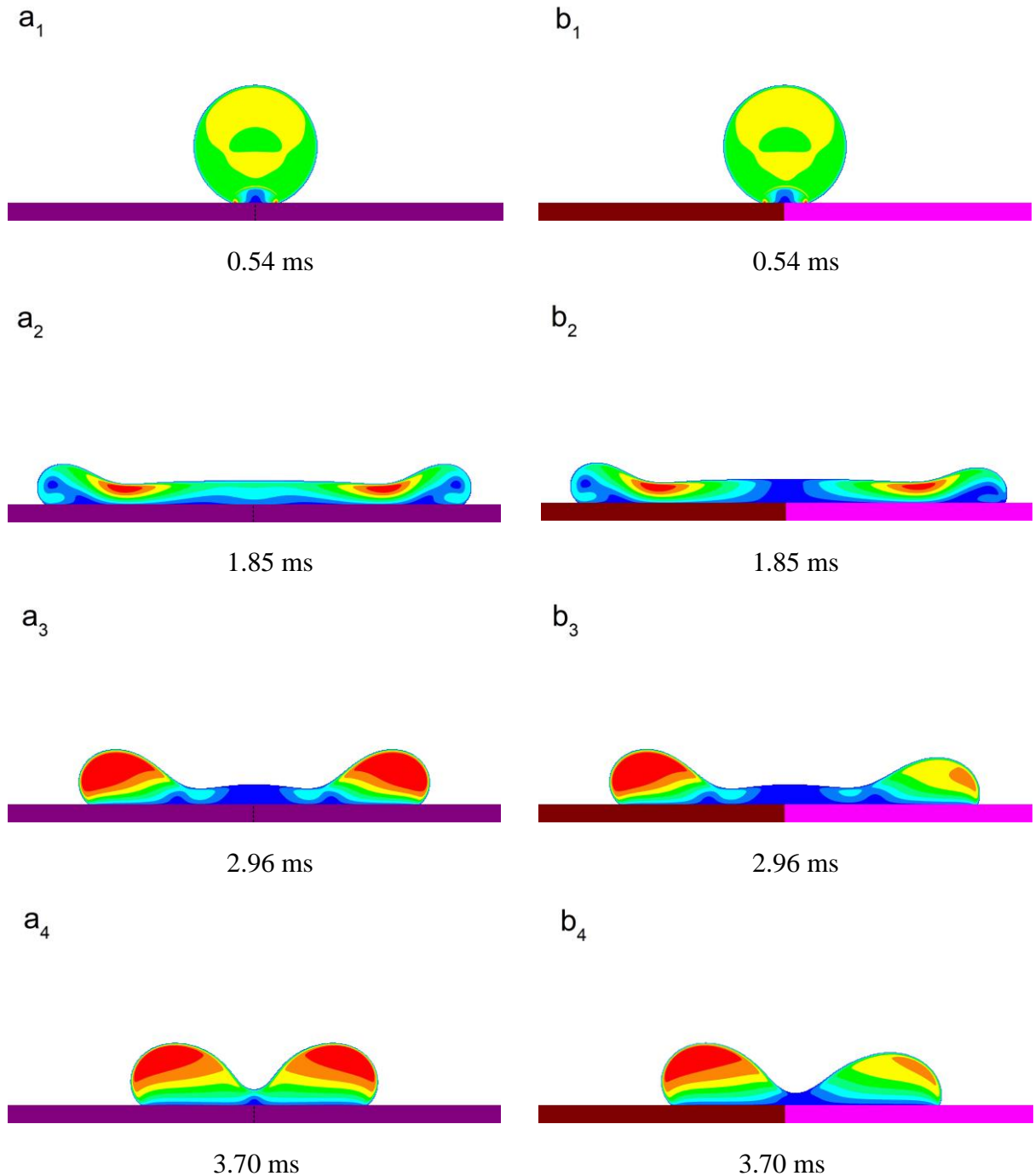

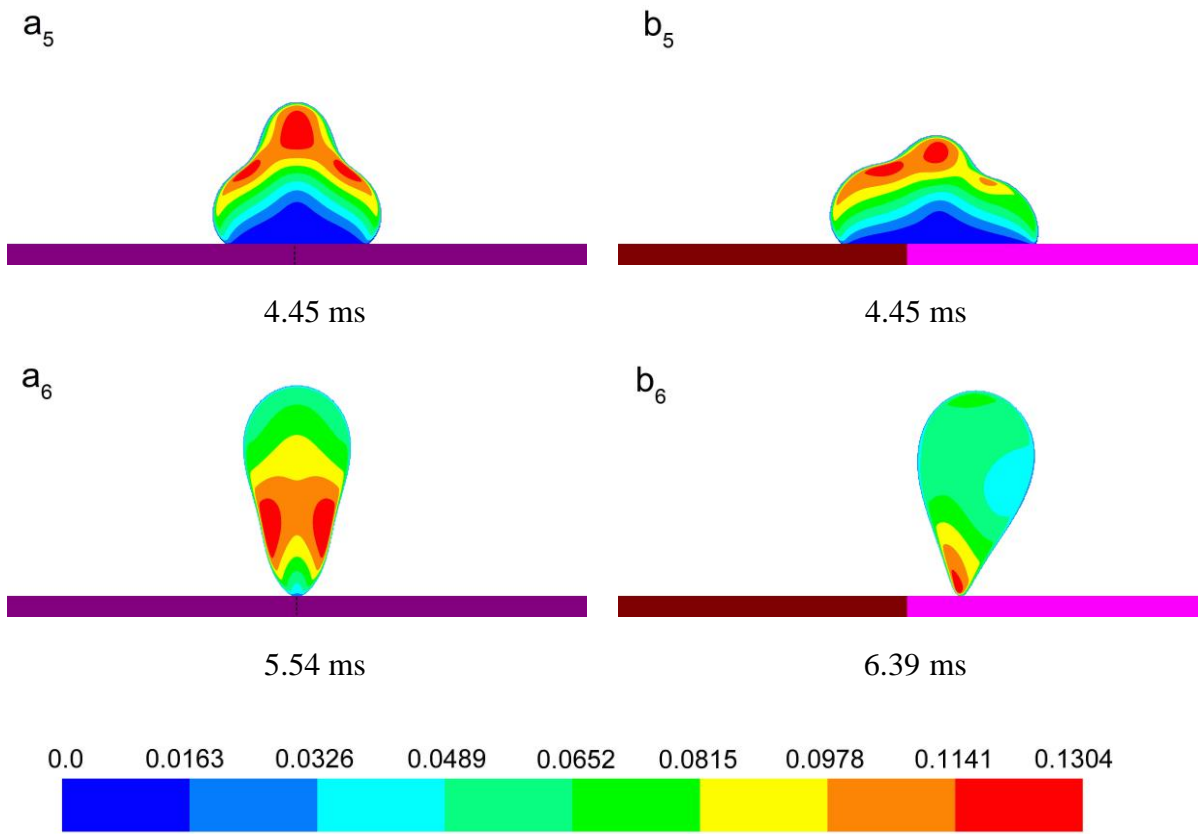

*Figure 7. Schematic diagram of the drop momentum mode. Among them, ($a_1$–$a_6$ ) is the change graph of the momentum modulus at different times when the surface contact angle is 150°, and ($b_1$–$b_6$) is the change graph of the momentum modulus of the drop when the angle of the heterogeneous surface is 150°–120°.*

On a heterogeneous surface, the typical rebound behavior of the drop is composed of spreading and retracting, and is an asymmetric process.[12, 32, 33] Figure 7($b_1$–$b_6$) is a graph showing the change of the drop momentum modulus when the angle of the heterogeneous surface is 150°–120°. When the drop hits the surface of the object, the drop firstly spreads rapidly until it reaches to a maximum lateral extension at the moment of 2.96 ms. However, during the retraction process, due to the heterogeneity of the surface, the hydrophobicity of the left side is greater than that of the right side, and the left side of the drop retracts more quickly, and finally the left side of the drop completely crosses the high hydrophobic surface and enters the low hydrophobic surface. Figure 7($b_6$) clearly shows that the drop is completely located on the low hydrophobic surface. It can also be observed from Figure 7($b_1$–$b_6$) that the momentum on the left of the drop is significantly greater than that on the right.

## 3.7. Mechanical analysis

Once the size and velocity of the drop remain constant, the behaviors of the drop impact and rebound are dependent on the surface properties. The vertical rebound of the drop is due to the high hydrophobicity of the solid surface, while the lateral movement is caused by the asymmetrical hydrophobicity of the two parts of the solid surface. In this section the force from the surface is discussed during the drop touching the surface. The two cases, a 150° homogeneous surface and a 150°−120° heterogeneous surface, are investigated.

Contact angle, which indicates the wettability of a solid surface, is a characteristic quantity in drop impacting. Ideally, the equilibrium contact angle of the sessile drop can be theoretically explained by Young's equation.[34] In the dynamic cases, in which the drop deforms and the three-phase contact line moves, the dynamic contact angle deviates from the equilibrium contact angle.[35-37] The dynamic interaction generates the unbalanced Young's force, which (per unit length) is expressed as[37]

$$F = \gamma(\cos\theta - \cos\theta_{eq}), \tag{0}$$

where $\gamma$ is the liquid−vapor interfacial tension, $\theta$ is the dynamic contact angle, and $\theta_{eq}$ is the equilibrium contact angle related to the surface properties. When the liquid−gas transition region is on a homogeneous surface, $\theta_{eq}$ is equal to the contact angle of the surface. If the surface is chemically heterogeneous and is composed by two kinds of surface components, the equilibrium contact angle of the heterogeneous surface is described by a modified Cassie-Baxter equation:[38]

$$\cos\theta_{eq} = r\varphi_d \cos\theta_{s1} + (1-\varphi_d)\cos\theta_{s2}, \tag{0}$$

where $\theta_{s1}$ and $\theta_{s2}$ are the intrinsic equilibrium contact angles of each surface component, $r$ is the "roughness" of the wetted surface ($r = 1$ for a chemically heterogeneous surface). The $\varphi_d$ and $1-\varphi_d$ are the areal ratios of the solid-liquid interface and the liquid-vapor interface, respectively. In the context of two-dimensional diffuse interface modeling, $\varphi_d$ and $1-\varphi_d$ respectively refer to the ratio of the length of the iso-density line on the surfaces with contact angles of $\theta_{s1}$ and $\theta_{s2}$ to the total length of the iso-density lines.[39] When the contact line is located at the junction of the high and low hydrophobic surface, the corresponding equilibrium contact angle can be estimated by Eq (22). When $\theta > \theta_{eq}$, a positive unbalanced Young's force would be generated at the right contact line, whereas when $\theta < \theta_{eq}$, a positive unbalanced Young's force would be generated at the left contact line.

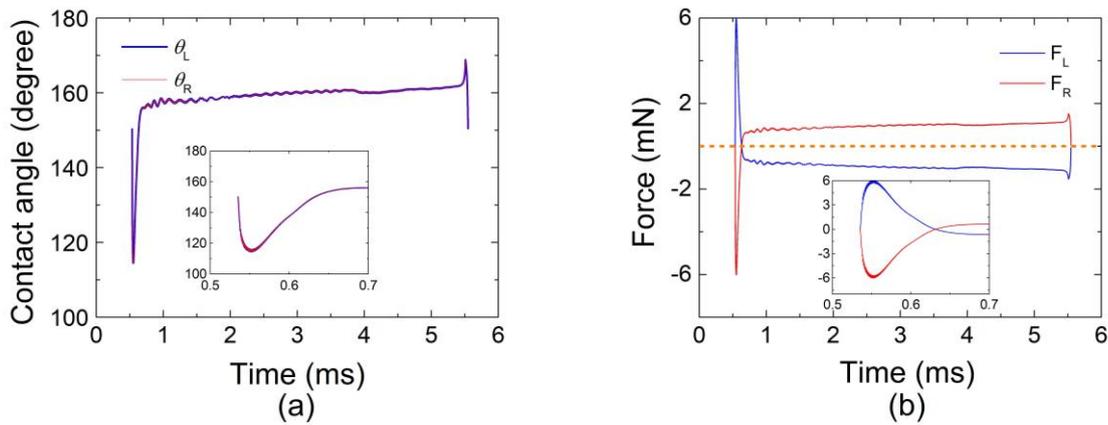

*Figure 8. Mechanical analysis of the drop impacting on the homogeneous surface. (a) Time evolution of the left and right dynamic contact angle of the drop. The inset shows the zoomed-in view between 0.5 ms and 0.7 ms. (b) Time evolution of unbalanced Young's force at the left and right contact lines. The inset shows the zoomed-in view between 0.5 ms and 0.7 ms.*

The variations of contact angle and unbalance force during the drop impacting on the homogeneous surface are displayed in Figure 8. Here, $\theta_L$ and $\theta_R$ represent the left and right dynamic contact angles respectively, $F_L$ and $F_R$, represent forces on the left and right contact point of the drop respectively. The results show that the left and right dynamic contact angles are identical when the drop hit a homogeneous surface. The dynamic contact angle changes sharply in the first 0.65 ms. This is because when the drop touches the solid surface, it is subjected to a great impact, and the drop produces a huge deformation as shown in Fig 7($a_1$–$a_2$). These also lead to the steep unbalanced Young's forces as shown clearly in Figure 8(b). The variation of dynamic contact angle tends to be smoothly following the impact disappear after 0.65 ms, and the contact line slides on the surface until 5.5 ms. The slight fluctuations are due to the high-frequency vibrations of the drop. Correspondingly, the unbalanced Young's force also tends to be stable from 0.65 ms to 5.5 ms. After 5.5 ms, the left and right three-phase contact regions are gradually approaching and influencing each other. Their deformations result in a steep vibration of the contact angle, and then the drop leaves the solid surface. It can be observed that during the whole process, the left and right forces (unbalanced Young's force) of the drop are symmetrical, so the entire drop (as shown in Figure 7($a_1$–$a_6$)) does not appear laterally inclined. The inset shows the zoomed-in view between 0.5 ms and 0.7 ms.

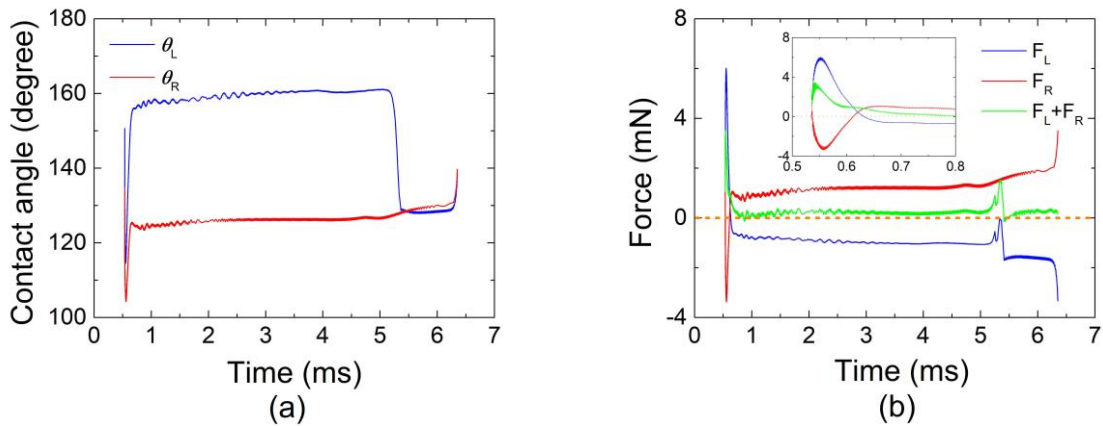

*Figure 9. Mechanical analysis of the drop impacting on the heterogeneous surface. (a) Time evolutions of the left and right dynamic contact angle of the drop. (b) Time evolution of the left and right unbalanced Young's force and its resultant force. The inset shows the zoomed-in view between 0.5 ms and 0.8 ms.*

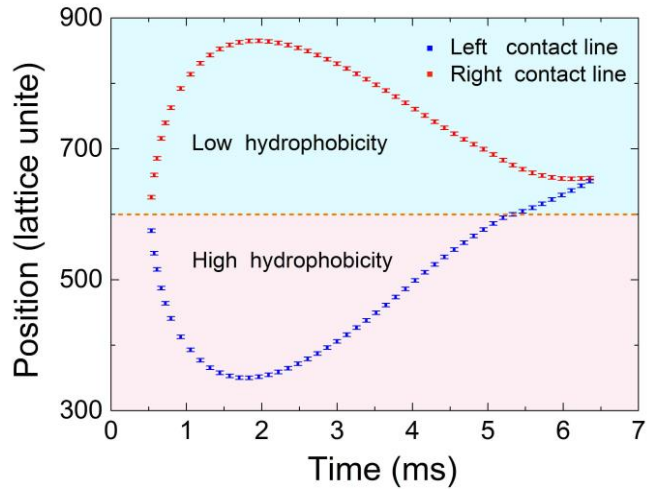

*Figure 10. Position evolution of the left and right contact line on the heterogeneous surface.*

When the surface morphology and chemical composition are heterogeneous, the dynamics of contact lines and contact angles are more complicated.[35, 40, 41] The variations of contact angle and unbalance force during the drop impacting on the heterogeneous surface are shown in Figure 9. Figure 9(a) shows the time evolution of the left and right dynamic contact angles on the heterogeneous surface. In the first 0.7 ms after the drop touch with the surface, also due to the role of great impact, the dynamic contact angle of the drop changes as drastically as the case on the homogeneous surface. Different from the homogeneous surface, the left and right contact lines fall on the heterogeneous surface with different wettability, and the left and right contact angles are obviously different. After that, the left dynamic contact angle of the drop is steady at about 160°, while right dynamic contact angle is near to 125°. At about 5.25 ms, the left dynamic contact angle of the drop dropped sharply because it just slips from the high hydrophobic surface to the low hydrophobic surface, as shown in Figure 10. After about 5.425 ms, the left side of the drop completely enters the low hydrophobic surface, and the left dynamic contact angle is slightly smaller than the right dynamic contact angle. Figure 9(b) shows the time evolution of the left and right unbalanced Young's force and its resultant force. The unbalanced Young's force changes drastically in the first 0.7 ms and then stabilizes. When the left contact line slips across the junction between the high and low hydrophobic surface, the unbalanced Young's force on the left side of the drop increases sharply, then decreases, and finally stabilizes. It can be observed that the left and right unbalanced Young's forces, are asymmetric and are in the opposite direction. Moreover, the unbalanced Young's force at right contact line is larger than that of at left contact line. Thus, the resultant force is always positive direction during the entire process from drop contacting to leaving the solid surface (see green line in Figure 9(b)). This causes that the drop deviates from the center of the substrate and finally lateral rebounds. Green line in figure 9(b) shows that it is at work. At the same time, this also explains the reason why the drop completely located in the low hydrophobic surface as shown in Figure 7($b_6$).

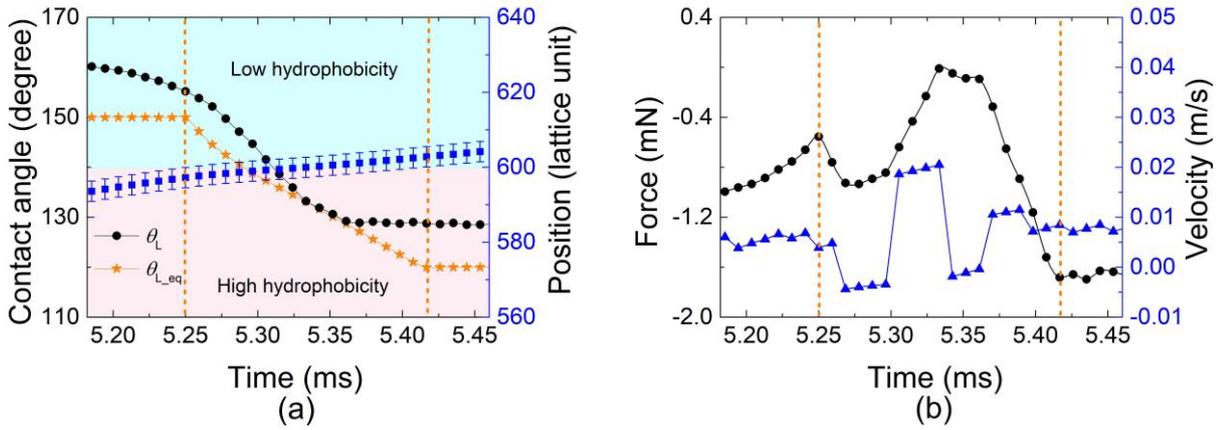

Figure 11. (*a*) Time evolution of the left dynamic contact angle (black line), the left equilibrium contact angle (orange line) and the position of the left contact line (blue line). (b) Time evolution of the unbalanced Young's force (black line) and the velocity of the left contact line (blue line). The two vertical orange dashed lines in Figure 11(a) and (b) indicate that the left contact line starts to enter and leave the junction of the low and high hydrophobic surfaces, respectively.

Furthermore, the process that the left contact line crosses over the junction of the low and high hydrophobic surface is analyzed in detail. Since the left parts have higher hydrophobicity than the right parts in the scheme II ~ IV, it is always the left contact line that crosses over the interface of two different surface components. Figure 11 shows the details of the time evolution of the left three-phase contact line, the left dynamic contact angle and the left equilibrium contact angle during the contact line moves from the high hydrophobic surface to the low hydrophobic surface. The equilibrium contact angle is calculated by Eq (22). The left contact line (blue line) of the drop starts to enter the junction of the low and high hydrophobic surfaces at about 5.25 ms as shown in Figure 11(a). The left dynamic contact angle and the left equilibrium contact angle continuously decrease, until 5.425 ms, the left contact line of the drop completely enters the low hydrophobic surface, and the left dynamic contact angle and the left equilibrium contact angle tend to be stable. Figure 11(b) is the time evolution of the unbalanced Young's force (black line) and velocity of the left contact line (orange line) when the left contact line of the drop enters the low hydrophobic surface from the high hydrophobic surface. It can be seen from Figure 11(b) that the unbalanced Young's force (black line) undergoes up-and-down fluctuations twice. The first time the left contact line position begins to reach the boundary of the junction of the low and high hydrophobic surfaces, it receives a certain resistance, and the unbalanced Young's force decreases, and the velocity of the left contact line reduces accordingly. As the left contact line continues to move into the junction of the low and high hydrophobic surfaces, the left dynamic contact angle and the left equilibrium contact angle continue to approach, and the unbalanced Young's force begins to increase. The velocity of the left contact line increases. When the left dynamic contact angle and the left equilibrium contact angle are equal, the unbalanced Young's force reaches zero, and the velocity of the left contact line reaches a peak of about 0.02 m/s. After that, the left dynamic contact angle deviates from the equilibrium contact angle, and the unbalanced Young's force and the velocity of the left contact line decrease. When the left contact line completely enters the low hydrophobic surface, the unbalanced Young's force and the velocity of the left contact line tend to stabilize.

## 4. Conclusion

In summary, we numerically investigate the rebounding mechanism of the drop impacting on the hydrophobic and chemically heterogeneous surface using the multiphase lattice Boltzmann method. It is found that when the drop impacting a homogeneous surface (Scheme I), the rebounding height and flying time increase with the increase of the surface hydrophobicity. When the drop impacts a chemically heterogeneous surface, the rebounding height and flying time also increase as the contact angle becomes larger. However, the lateral rebounding distance is dependent on the heterogeneous condition of the surface. When the right part is fixed low hydrophobicity and the hydrophobicity of the left part grows gradually (Scheme II), the lateral distance is farther and farther. When the left part is fixed with high hydrophobicity and the hydrophobicity of the right part gradually increases from low to high (Scheme III), the lateral distance first increases and then decreases. This is also observed when the surface has a fixed difference of the left and right contact angles (Scheme IV). Consequently, both Scheme III and IV have the maximum lateral bouncing distance, which occurs around 130 ° contact angle in Scheme III and 160º − 140 ° combination in Scheme IV. It tells that the proper hydrophobic combination is the key to obtain the most efficient lateral transport of the drop by using the chemical heterogeneity of the surface. Furthermore, analyzing the internal momentum distribution of the drop reveals the internal movement of the drop and deepens the understanding of the mechanism of the lateral rebound. Finally, based on the dynamic contact angle measurement and the unbalanced Young's force, we have made the mechanical analyses of the dynamic behavior. It is shown that the main reason for the lateral rebounding is the asymmetry of the left and right unbalanced Young's force on the heterogeneous surface. This left-right asymmetry force exists in the whole process that the drop contacts the heterogeneous surface including touching, spreading, shrinking, and even starting to leave, and finally causes the drop to move away from the center of the substrate to the low hydrophobic surface. This study can be expected to be helpful for accurate control of drop rebounding, fluid-solid driving, or energy harvest and storage.

**Acknowledgment**

This work was supported by the National Natural Science Foundation of China (Grant Nos. 11862003, 81860635), the Key Project of Guangxi Natural Science Foundation (Grant No. 2017GXNSFDA198038), the Graduate Innovation Program of Guangxi Normal University (Grant No. XYCSZ2020069), Guangxi "Bagui Scholar" Teams for Innovation and Research Project, Guangxi Collaborative Innovation Center of Multi-source Information Integration and Intelligent Processing.

# References

[1] JOSSERAND C, THORODDSEN S T. Drop Impact on a Solid Surface [J]. Annu Rev Fluid Mech, 2016, 48(1): 365-91.
[2] JOUNG Y S, BUIE C R. Aerosol generation by raindrop impact on soil [J]. Nature Communications, 2015, 6(1): 6083.
[3] RYKACZEWSKI K, PAXSON A T, STAYMATES M, et al. Dropwise Condensation of Low Surface Tension Fluids on Omniphobic Surfaces [J]. Scientific Reports, 2014, 4(1): 4158.
[4] FURUTA T, SAKAI M, ISOBE T, et al. Evaporation Behavior of Microliter- and Sub-nanoliter-Scale Water Droplets on Two Different Fluoroalkylsilane Coatings [J]. Langmuir, 2009, 25(20): 11998-2001.
[5] SINGH M, HAVERINEN H M, DHAGAT P, et al. Inkjet Printing-Process and Its Applications [J]. Adv Mater, 2010, 22(6): 673-85.
[6] MASSINON M, LEBEAU F. Experimental method for the assessment of agricultural spray retention based on high-speed imaging of drop impact on a synthetic superhydrophobic surface [J]. Biosystems Engineering, 2012, 112(1): 56-64.
[7] SONG M, JU J, LUO S, et al. Controlling liquid splash on superhydrophobic surfaces by a vesicle surfactant [J]. Science Advances, 2017, 3(3): e1602188.
[8] BIRD J C, DHIMAN R, KWON H M, et al. Reducing the contact time of a bouncing drop [J]. Nature, 2013, 503(7476): 385-8.
[9] MALOUIN B A, KORATKAR N, HIRSA A, et al. Directed rebounding of droplets by microscale surface roughness gradients [J]. Appl Phys Lett, 2010, 96(23): 234103.
[10] ANTONINI C, AMIRFAZLI A, MARENGO M. Drop impact and wettability: From hydrophilic to superhydrophobic surfaces [J]. Phys Fluids, 2012, 24(10): 102104-17.
[11] YAO C, GARVIN T P, ALVARADO J L, et al. Droplet contact angle behavior on a hybrid surface with hydrophobic and hydrophilic properties [J]. Appl Phys Lett, 2012, 101(11): 111605.
[12] WEISENSEE P B, TIAN J, MILJKOVIC N, et al. Water droplet impact on elastic superhydrophobic surfaces [J]. Scientific Reports, 2016, 6(1): 30328-9.
[13] LI H, FANG W, LI Y, et al. Spontaneous droplets gyrating via asymmetric self-splitting on heterogeneous surfaces [J]. Nature Communications, 2019, 10(1): 1-6.
[14] HAO J, LU J, LEE L, et al. Droplet Splashing on an Inclined Surface [J]. Phys Rev Lett, 2019, 122(5): 054501.
[15] HAN J, KIM W, BAE C, et al. Contact time on curved superhydrophobic surfaces [J]. Phys Rev E, 2020, 101(4): 043108.
[16] LU Y, SHEN Y, TAO J, et al. Droplet directional movement on the homogeneously structured superhydrophobic surface with the gradient non-wettability [J]. Langmuir, 2020, 36(4): 880-8.
[17] ZHANG B, LEI Q, WANG Z, et al. Droplets Can Rebound toward Both Directions on Textured Surfaces with a Wettability Gradient [J]. Langmuir, 2015, 32(1): 346-51.
[18] YUAN W-Z, ZHANG L-Z. Lattice Boltzmann simulation of droplets impacting on superhydrophobic surfaces with randomly distributed rough structures [J]. Langmuir, 2017, 33(3): 820-9.
[19] PENG B, WANG S, LAN Z, et al. Analysis of droplet jumping phenomenon with lattice Boltzmann simulation of droplet coalescence [J]. Appl Phys Lett, 2013, 102(15):151601.

[20] RAMAN K A, JAIMAN R K, LEE T-S, et al. Lattice Boltzmann simulations of droplet impact onto surfaces with varying wettabilities [J]. Int J Heat Mass Transfer, 2016, 95:336-54.

[21] HOSEINPOUR B, SARRESHTEHDARI A, ASHORYNEJAD H R. Lattice Boltzmann simulation of dynamics of droplet impact on inclined walls [J]. Int J Mod Phys C, 2019, 30(8):

[22] WEN B, ZHANG C, TU Y, et al. Galilean invariant fluid-solid interfacial dynamics in lattice Boltzmann simulations [J]. J Comput Phys, 2014, 266:161-70.

[23] WEN B, HUANG B, QIN Z, et al. Contact angle measurement in lattice Boltzmann method [J]. Comput Math Appl, 2018, 76(7): 1686-98.

[24] WEN B, ZHOU X, HE B, et al. Chemical-potential-based lattice Boltzmann method for nonideal fluids [J]. Physical review E, 2017, 95(6-1): 063305.

[25] KUPERSHTOKH A L, MEDVEDEV D A, KARPOV D I. On equations of state in a lattice Boltzmann method [J]. Comput Math Appl, 2009, 58(5): 965-74.

[26] ROWLINSON B J S, WIDOM B. Molecular theory of capillarity [M]. Clarendon*, 1982.

[27] SWIFT M R, OSBORN W R, YEOMANS J M. Lattice Boltzmann simulation of nonideal fluids [J]. Phys Rev Lett, 1995, 75(5): 830-3.

[28] WEN B, QIN Z, ZHANG C, et al. Thermodynamic-consistent lattice Boltzmann model for nonideal fluids [J]. EPL (Europhysics Letters), 2015, 112(4):44002.

[29] JAMET D, TORRES D, BRACKBILL J U. On the Theory and Computation of Surface Tension: The Elimination of Parasitic Currents through Energy Conservation in the Second-Gradient Method [J]. J Comput Phys, 2002, 182(1): 262-76.

[30] ZHENG H W, SHU C, CHEW Y T. A lattice Boltzmann model for multiphase flows with large density ratio [J]. J Comput Phys, 2006, 218(1): 353-71.

[31] WEN B, ZHAO L, QIU W, et al. Chemical-potential multiphase lattice Boltzmann method with superlarge density ratios [J]. Physical review E, 2020, 102(1-1): 013303.

[32] LI D, FENG S, XING Y, et al. Directional bouncing of droplets on oblique two-tier conical structures [J]. RSC Advances, 2017, 7(57): 35771-5.

[33] LI J, GUO Z. Spontaneous directional transportations of water droplets on surfaces driven by gradient structures [J]. Nanoscale, 2018, 10(29): 13814-31.

[34] YOUNG T. An Essay on the Cohesion of Fluids [J]. Philosophical Transactions of the Royal Society, 1805, 95.

[35] EXTRAND C W. Contact Angles and Hysteresis on Surfaces with Chemically Heterogeneous Islands [J]. Langmuir, 2003, 19(9): 3793-6.

[36] ELSHERBINI A, JACOBI A M. Liquid drops on vertical and inclined surfaces: I. An experimental study of drop geometry [J]. J Colloid Interface Sci, 2004, 273(2): 556-65.

[37] SUI Y, DING H, SPELT P D M. Numerical Simulations of Flows with Moving Contact Lines [J]. Annu Rev Fluid Mech, 2014, 46(1): 97-119.

[38] CHOI W, TUTEJA A, MABRY J M, et al. A modified Cassie-Baxter relationship to explain contact angle hysteresis and anisotropy on non-wetting textured surfaces [J]. J Colloid Interface Sci, 2009, 339(1): 208-16.

[39] LI Q, ZHOU P, YAN H J. Pinning–Depinning Mechanism of the Contact Line during Evaporation on Chemically Patterned Surfaces: A Lattice Boltzmann Study [J]. Langmuir, 2016, 32(37): 9389-96.

[40] JANSEN H P, BLIZNYUK O, KOOIJ E S, et al. Simulating Anisotropic Droplet Shapes on Chemically Striped Patterned Surfaces [J]. Langmuir, 2012, 28(1): 499-505.

[41] VARAGNOLO S, FERRARO D, FANTINEL P, et al. Stick-slip sliding of water drops on chemically heterogeneous surfaces [J]. Phys Rev Lett, 2013, 111(6): 066101.

TOC Graphic:

we used the multiphase lattice Boltzmann method driven by a chemical potential to analyze the rebounding behavior of the drop. The trajectory changes, rebounding height, lateral movement distance, and flying time of the droplets after impacting the heterogeneous surface are analyzed in detail.

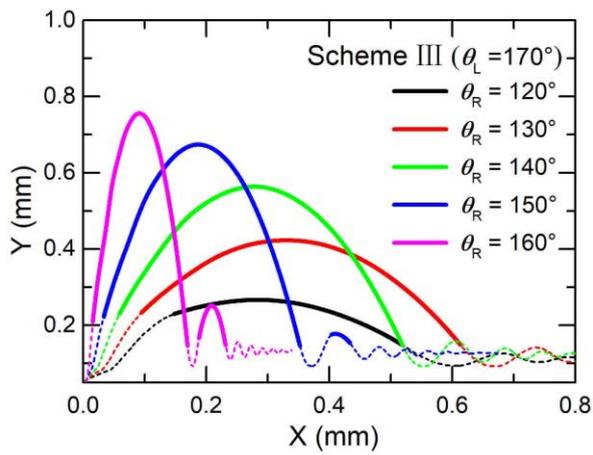
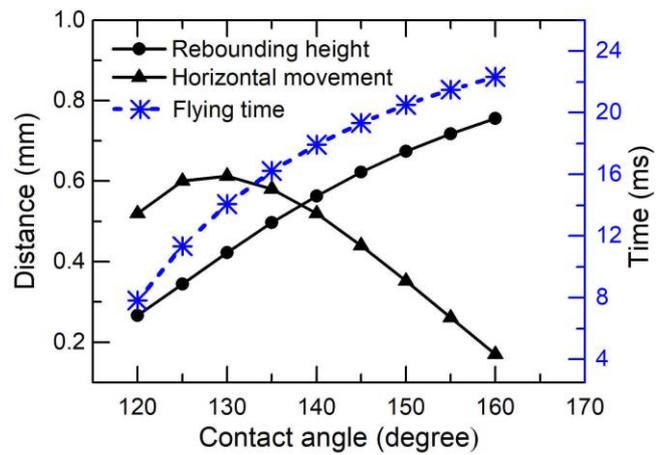